\def\Ar{\rightarrow}
\def\bar{\overline}

\def\d{\delta}
\def\a{\alpha}
\def\b{\beta}
\def\n{\nu}
\def\m{\mu}

\def\e{\epsilon}

\def\th{\theta}

\def\bar{\overline}
\def\l{\lambda}

\def\eV{{\rm eV}}
\documentstyle [12pt,epsf]{article}
\begin{document}
\baselineskip=24pt
\setcounter{page}{1}
\thispagestyle{empty}
\topskip 2.5  cm
\begin{flushright}
\begin{tabular}{c c}
& EHU-00-01, January 2000
\end{tabular}
\end{flushright}
\vspace{1 cm}
\centerline{\LARGE \bf Large Mixing Angle MSW Solution}
\vskip 0.5 cm
\centerline{\LARGE\bf  in $S_3$ Flavor Symmetry}
\vskip 1.5 cm
\centerline{{\large \bf Morimitsu TANIMOTO}
  \footnote{E-mail address: tanimoto@edserv.ed.ehime-u.ac.jp}
   }
\vskip 0.8 cm
 \centerline{ \it{Science Education Laboratory, Ehime University, 
 790-8577 Matsuyama, JAPAN}}
\vskip 3 cm
\centerline{\bf ABSTRACT}\par
\vskip 0.5 cm
  We have investigated  phenomenological implications on the neutrino
 flavor mixings in the $S_{3L}\times S_{3R}$ symmetric mass matrices
 including symmetry breaking terms. 
  We have shown how to get the large mixing angle MSW solution, 
   $\sin^2 2\th_\odot=0.65\sim 0.97$ and
   $\Delta m_{\odot}^2= 10^{-5}\sim 10^{-4}\eV^2$, in this model.
  It is found that the structure of the lepton mass matrix 
  in our model is stable against radiative corrections 
  although the model leads to nearly degenerate neutrinos.
  
  \vskip 0.5 cm
 \newpage
\topskip 0. cm

  Recent Super-Kamiokande data of atmospheric neutrinos \cite{SKam} have provided 
  a more solid evidence of the neutrino oscillation, which corresponds to 
  the nearly maximal neutrino flavor mixing.
  The observed solar neutrino deficit 
  is also an indication of a different sort of the neutrino oscillation
  \cite{SKamsolar}. 
   For the solar neutrino problem, four solutions are still allowed.
   Those are large mixing angle (LMA) MSW, small mixing angle (SMA) MSW \cite{MSW},
   vacuum oscillation (VO) and low $\Delta m^2$ (LOW) solutions \cite{BKS}.
	
	Those data give constraints on the structure of the 
	lepton mass matrices in the three family model \cite{Texture,FX,FTY}, 
	which may suggest the some flavor symmetry \cite{FS,O3}.
   There is a typical texture of the lepton mass matrix with
	  the nearly maximal mixing of flavors, which
	 is derived from the symmetry of the lepton flavor democracy \cite{FX}, 
	 or  from the $S_{3L}\times S_{3R}$ symmetry 
	 of the left-handed Majorana neutrino mass matrix \cite{FTY}. 
	 This texture have given a prediction for the neutrino mixing 
	  $\sin^2 2\th_{\rm atm}=8/9$. The mixing for the solar neutrino
	  depends on the symmetry breaking pattern of the flavor such as
	  $\sin^2 2\th_\odot= 1$ or $\ll 1$.
	However, the LMA-MSW solution, $\sin^2 2\th_\odot=0.65\sim 0.97$ and
	$\Delta m_{\odot}^2= 10^{-5}\sim 10^{-4}\eV^2$,  
	  has not been obtained in the previous works \cite{FX,FTY}.
	  
   In this paper, we study how to get the LMA-MSW solution
   in the $S_{3L}\times S_{3R}$ symmetric mass matrices 
   including symmetry breaking terms.
  Furthermore, we discuss the stability of the neutrino mass matrix against radiative 
  corrections since the model predicts nearly degenerate neutrinos.

 We assume that oscillations need only account for 
 the solar and atmospheric neutrino data.
 Since the result of LSND \cite{LSND} awaits confirmation by KARMEN experiment
 \cite{KARMEN},
  we do not take into consideration the LSND data in this paper.
  Our starting point as to the neutrino mixing is
  the large $\nu_\mu \Ar \nu_\tau$ oscillation of  atmospheric neutrinos  with 
 $\Delta m^2_{\rm atm}=  (2\sim 6)\times  10^{-3} \eV^2$ and 
 $\sin^2 2\th_{\rm atm} \geq 0.84$,
 which is derived from the recent data of the atmospheric neutrino deficit 
 at Super-Kamiokande \cite{SKam}. 
 The mass difference scales of the solar neutrinos
  are $\Delta m_{\odot}^2= 10^{-10}\sim 10^{-4}\eV^2$ depending on the
  four solutions \cite{BKS}.


The texture of the charged lepton mass matrix
was presented based on the $S_{3L}\times S_{3R}$ symmetry
  as follows \cite{FX,FTY,Koide}:
\begin{equation}
M_\ell= {c_\ell \over 3}
             \left (\matrix{1 & 1 & 1 \cr
			                1 & 1 & 1 \cr
                            1 & 1 & 1 \cr  } \right)  + M_\ell^{(c)}\ ,
\label{MC}
\end{equation}
\noindent where the second matrix is the flavor symmetry breaking one.
The unitary matrix $V_\ell$, which diagonalizes the mass matrix $M_\ell$,
 is given as $V_\ell=F L$, where
\begin{equation}
F=\left( \matrix{1/\sqrt 2 & 1/\sqrt 6 & 1/\sqrt 3 \cr
                   -1/\sqrt 2 & 1/\sqrt 6 & 1/\sqrt 3 \cr
                           0 & -2/\sqrt 6 & 1/\sqrt 3 \cr } \right) 
\label{F} 
\end{equation}										 
\noindent diagonalizes the democratic matrix and $L$ depends on
the mass correction term $M_\ell^{(c)}$.

 Let us turn to the neutrino sector. 
 The neutrino mass matrix is different from the democratic one 
 if they are Majorana particles.
The $S_{3L}$ symmetric mass term is given as follows:
\begin{equation}
   {c_\n}  \left( \matrix{1 & 0 & 0 \cr
                            0 & 1 & 0 \cr
                            0 & 0 & 1 \cr  } \right)
	+ {c_\n} r \left( \matrix{1 & 1 & 1 \cr
                            1 & 1 & 1 \cr
                            1 & 1 & 1 \cr  } \right) \ , 
  \end{equation}
\noindent
where $c_\n$ and $r$ are arbitrary parameters.
The eigenvalues of this matrix are easily obtained by using the orthogonal matrix
 $F$ in eq.(\ref{F}) as 
$c_\n(1, \ 1, \ 1+3r)$, which means that there are at least two degenerate masses
 in the $S_{3L}$ symmetric Majorana mass matrix \cite{FTY,KK,Lisbon}.
 
 The simplest breaking terms of the $S_{3L}$ symmetry are added in (3,3)
 and (2,2) entries.  Therefore, the neutrino mass matrix is written as
 \begin{equation}
 M_\n= {c_\n}  \left( \matrix{1 + r & r & r \cr
                              r & 1+r+\e & r \cr
                              r & r & 1+r+\d \cr  } \right) \ ,
 \label{Mass}
 \end{equation}
 \noindent
 in terms of small breaking parameters $\e$ and $\d$.
 In order to explain both solar and atmospheric neutrinos in this mass matrix,
  $r\ll 1$ should be satisfied. 
  In other words, three neutrinos should be nearly 
  degenerate.\footnote{$r=-2/3$ also gives nearly degenerate neutrinos \cite{KK}.
   However, solar and atmospheric neutrinos are not explained by simple 
  breaking terms in eq.(\ref{Mass}).} 
  However, there is no reason why $r$ is very small in this framework.
  In order to answer this question,  we need a higher symmetry 
  of  flavors such as  the $O_{3L}\times O_{3R}$ model \cite{O3}.
  We do not address this problem in this paper.

  We start with discussing the simple case of $\e=0$
  and $\d\gg r$, in which  the $S_{2L}$ symmetry
  is preserved  but  the $S_{3L}$ symmetry is broken.
  Mass eigenvalues are given as
\begin{equation}
  m_1 = 1 \ , \quad  m_2 \simeq 1 + 2 r\ , \quad  m_3 \simeq 1 + r + \d \ ,
\end{equation}
\noindent in the $c_\n$ unit.
We easily obtain $\Delta m_{\rm atm}^2=\Delta m_{32}^2\simeq 2c_\n^2 \d$ and 
  $\Delta m_{\odot}^2=\Delta m_{21}^2\simeq 4 c_\n^2 r$.
The neutrino mass matrix is diagonalized by the orthogonal matrix $U_\n$ such as
$ U_\n^T M_\n U_\n$, where

\begin{equation}
U_\n \simeq \left (\matrix{\frac{1}{\sqrt{2}} & \frac{1}{\sqrt{2}} & \frac{r}{\d} \cr
			         -\frac{1}{\sqrt{2}} & \frac{1}{\sqrt{2}} & \frac{r}{\d} \cr
                            0 & -\sqrt{2}\frac{r}{\d} & 1 \cr  } \right)   ,
\label{Mix1}
\end{equation}

\noindent
 in which the first and second family mixes maximally due to the $S_{2L}$ symmetry.
This maximal mixing is completely canceled out by the charged lepton
 sector in the neutrino mixing matrix (MNS mixing matrix) $U_{\a i}$ \cite{MNS},
  which is determined 
  by the product of $V_\ell^\dagger$ and $U_\n$  as follows: 
   
 \begin{equation}
  U=V^{\dagger}_\ell U_\n=L^\dagger F^T U_\n \simeq \left (
  \matrix{1 & \frac{1}{\sqrt{3}} L_{21} & -\sqrt{\frac{2}{3}} L_{21}\cr
			  L_{12}& \frac{1}{\sqrt{3}}(1+2 \frac{r}{\d}+\sqrt{2}L_{32})
  & -\sqrt{\frac{2}{3}}(1- \frac{r}{\d}+\frac{1}{\sqrt{2}}L_{32}) \cr
   L_{13} &\sqrt{\frac{2}{3}}(1- \frac{r}{\d}+\frac{1}{\sqrt{2}}L_{23}) 
   & \frac{1}{\sqrt{3}}(1+2 \frac{r}{\d}-\sqrt{2}L_{23}) \cr  } \right) \ ,
   \label{MNS1}
  \end{equation}
 
  \noindent
  where $L_{ij}$ are components of the correction matrix $L$ 
  in the charged lepton sector. We take  $L_{ii}\simeq 1$$(i=1,2,3)$
  and  $L_{31}\ll L_{21}\ll 1$ like mixings in the quark sector. 
  The CP violating phase is also neglected.
  This case corresponds to the SMA-MSW solution of the solar neutrino.
  In this MNS mixing matrix, we have:
\begin{equation}
  U_{e3}\simeq  -\sqrt{2} U_{e2} \ ,
\end{equation}
\noindent
which means that $U_{e3}$ is predicted if the solar neurino data will be confirmed
in the future.  The long baseline (LBL) experiments provide an important test of 
the model since  the oscillation of $\n_\mu \Ar \n_e$ is predicted as follows:
\begin{equation}
     P(\n_\mu \Ar \n_e)\simeq \frac{4}{3}\sin^2 2\th_\odot 
      \sin^2 \frac{\Delta m_{31}^2 L}{4 E} \ .
\end{equation}
\noindent
Putting $\sin^2 2\th_\odot$ of the SMA-MSW solution \cite{BKS}, we obtain
  $P(\n_\mu \Ar \n_e)=10^{-3}\sim 10^{-2}$ in the relevant LBL experiment.
These results with the SMA-MSW solution of the solar neutrino
 are maintained as far as  $\e \ll r$.
 
  Let us consider the case of $\e \not= 0$
  with $\d\gg \e \simeq r$, in which  $S_{3L}$ symmetry is completely broken.
  Then neutrino mass eigenvalues are given as
\begin{equation}
  m_1 \simeq 1 + \frac{1}{2}\e+ r - \frac{1}{2}\sqrt{\e^2+4 r^2} \ , \quad 
  m_2 \simeq 1 + \frac{1}{2}\e+ r + \frac{1}{2}\sqrt{\e^2+4 r^2}\ , 
  \quad   m_3 \simeq 1 + r + \d ,
\end{equation}
\noindent in the $c_\n$ unit.
Then we have
\begin{equation}
   \Delta m_{32}^2 \simeq 2 c_\n^2 \d \ , \qquad 
   \Delta m_{21}^2 \simeq 2 c_\n^2 \sqrt{\e^2+4 r^2} \ .
 \label{dm2}
\end{equation}
\noindent
The orthogonal matrix $U_\n$ is given as
 \begin{equation}
U_\n \simeq \left (\matrix{t & \sqrt{1-t^2} & \frac{r}{\d} \cr
			         -\sqrt{1-t^2} & t & \frac{r}{\d-\e} \cr
  \frac{r}{\d}(\sqrt{1-t^2}-t) & -\frac{r}{\d-\e}(t+\sqrt{1-t^2}) & 1 \cr}\right)\ ,
\label{Mix2}
\end{equation}
\noindent where
\begin{equation}
  t^2=\frac{1}{2}+\frac{1}{2}\frac{\e}{\sqrt{\e^2+4r^2}} \ . 
\end{equation}
 
 In order to find the structure of the MNS matrix $U_{\a i}$, 
 we show $F^T U_\n$ as follows:
 
  \begin{equation}
   F^T U_\n \simeq 
   \left (\matrix{\frac{1}{\sqrt{2}}(t+\sqrt{1-t^2})
 &\frac{1}{\sqrt{2}}(\sqrt{1-t^2}-t) & -\frac{1}{\sqrt{2}}\frac{\e r}{\d(\d-\e)} \cr
	\frac{1}{\sqrt{6}}(t-\sqrt{1-t^2})(1+\frac{2r}{\d})
	 & \frac{1}{\sqrt{6}}(t+\sqrt{1-t^2})(1+\frac{2r}{\d-\e})
	 & -\frac{2}{\sqrt{6}}(1-\frac{r}{\d}) \cr
  \frac{1}{\sqrt{3}}(t-\sqrt{1-t^2})(1-\frac{r}{\d})
   &  \frac{1}{\sqrt{3}}(t+\sqrt{1-t^2})(1-\frac{r}{\d-\e}) 
   & \frac{1}{\sqrt{3}}(1+\frac{2r}{\d}) \cr}\right)\ .
  \end{equation}
  
  \noindent
  The mixing angle between the first and second flavor depends on $t$, which
  is determined by $r/\e$. 
   It becomes the maximal angle in the case of $t=1$ ($r/\e=0$) and
   the minimal  one in the case of $t=1/\sqrt{2}$ ($\e/r=0$).
   It is emphasized that the relevant value of  $r/\e$ leads easily to 
    $\sin^2 2\th_\odot=0.65\sim 0.97$, which corresponds to the LMA-MSW solution.
    The case of $t=1/\sqrt{2}$ may correspond rather to the VO solution.
    
   In order to get the MNS mixing matrix $U_{\a i}$,  the correction
     matrix $L^\dagger$ in the charged lepton sector should be multiplied such as 
    $L^\dagger F^T U_\n$.  Then we obtain:
    
  \begin{eqnarray}
 &&U_{e1}\simeq\frac{1}{\sqrt{2}}(t+\sqrt{1-t^2})
        +\frac{1}{\sqrt{6}}(t-\sqrt{1-t^2})L_{21} \ , \nonumber\\
 &&U_{e2}\simeq\frac{1}{\sqrt{2}}(\sqrt{1-t^2}-t)
        +\frac{1}{\sqrt{6}}(t+\sqrt{1-t^2})L_{21} \ , \nonumber\\
 &&U_{e3}\simeq  -\frac{2}{\sqrt{6}}(1-\frac{r}{\d})L_{21}  \ , \nonumber\\ 
 &&U_{\mu1}\simeq\frac{1}{\sqrt{6}}(t-\sqrt{1-t^2})(1+\frac{2r}{\d})
        +\frac{1}{\sqrt{2}}(t+\sqrt{1-t^2})L_{12} \ , \nonumber\\ 
 &&U_{\mu2}\simeq\frac{1}{\sqrt{6}}(t+\sqrt{1-t^2})(1+\frac{2r}{\d})
        +\frac{1}{\sqrt{2}}(\sqrt{1-t^2}-t)L_{12} \ , \nonumber\\ 
 &&U_{\mu3}\simeq  -\frac{1}{\sqrt{6}}(2-\frac{2r}{\d}-\sqrt{2}L_{32})\ ,  
               \label{LMIX} \\ 
 &&U_{\tau1}\simeq\frac{1}{\sqrt{3}}(t-\sqrt{1-t^2})
     (1-\frac{r}{\d}+\frac{1}{\sqrt{2}}L_{23})\ ,\nonumber\\
 &&U_{\tau2}\simeq\frac{1}{\sqrt{3}}(t+\sqrt{1-t^2})
     (1-\frac{r}{\d}+\frac{1}{\sqrt{2}}L_{23})\ ,\nonumber\\
 &&U_{\tau3}\simeq  \frac{1}{\sqrt{3}}(1+\frac{2r}{\d}-\sqrt{2}L_{23})\ ,\nonumber 
  \end{eqnarray}  
 \noindent
 where $L_{ii}\simeq 1$$(i=1,2,3)$ are taken and $L_{31}, L_{13}$ are neglected. 
 The CP violating phase is also neglected. 
$U_{e3}$ depends on $L_{21}$, which is determined by $M_\ell^{(c)}$ in eq.(\ref{MC}).
 The MNS mixings in  eqs.(\ref{LMIX}) agree with the numerical one
 (without any approximations)  within a few percent error.
  
  We should carefully discuss the stability of our results against radiative 
  corrections since the model predicts nearly degenerate neutrinos.
  When the texture of the mass matrix is given 
  at the $S_{3L}\times S_{3R}$ symmetry energy scale, radiative corrections are not
  negligible at the electoroweak (EW) scale.
  The runnings of the neutrino masses and mixings have been studied by using
  the renormalization group equations (RGE's) \cite{RGE1,RGE2,Haba}.
  
  Let us consider the basis, in which the mass matrix of the charged leptons
  is diagonal.  The neutrino mass matrix in eq.(\ref{Mass}) is transformed into
  $V_\ell^\dagger M_\n V_\ell$.
  Taking $V_\ell\simeq F$ because of $L$ being close to the unit matrix,
    we obtain the  mass matrix at the high energy scale:
 \begin{equation}
 F^T M_\n F=\bar M_\n= {c_\n}  
 \left(\matrix{1 + \frac{\e}{2} & -\frac{\e}{2\sqrt{3}} & -\frac{1}{\sqrt{6}}\e \cr
  -\frac{\e}{2\sqrt{3}}& 1+\frac{1}{6}\e+ \frac{2}{3}\d 
          & \frac{\sqrt{2}}{6}\e -\frac{\sqrt{2}}{3}\d \cr
  -\frac{1}{\sqrt{6}}\e & \frac{\sqrt{2}}{6}\e -\frac{\sqrt{2}}{3}\d 
  & 1+ \frac{1}{3} \e +  \frac{1}{3} \d + 3 r \cr  } \right) \ .
 \label{Massnew}
 \end{equation}
 \noindent
 The radiatively corrected mass matrix in the MSSM at the EW scale is given 
 as $R_G \bar M_\n R_G$, where $R_G$ is given by RGE's \cite{Haba} as
  \begin{equation}
 R_G\simeq \left(\matrix{1+ \eta_e & 0 & 0 \cr 0 & 1+\eta_\m & 0 \cr 0 & 0 & 1\cr}
  \right)\ ,
  \label{RGE}
  \end{equation}
  \noindent
  where $\eta_e$ and $\eta_\m$ are 
   \begin{equation}
  \eta_i = 1- \sqrt{\frac{I_i}{I_\tau}} \  (i=e,\ \mu) \ ,
  \end{equation}
  \noindent
  with
\begin{equation}
I_i\equiv\exp {\left (\frac{1}{8\pi^2}\int\limits_{\ln{(Mz)}}^{\ln{(M_R)}} y_i^2\ dt
      \right )} \ .
\end{equation}
Here $y_i\ (i=e,\ \mu)$ are Yukawa couplings and 
the $M_R$ scale is taken as the $S_{3L}\times S_{3R}$ symmetry energy scale.
We transform back  this neutrino mass matrix $R_G \bar M_\n R_G$ into the basis where the charged lepton mass matrix is the democratic one at the EW scale:
 \begin{equation}
   F R_G \bar M_\n R_G F^T \simeq {c_\n}\left( \matrix{1+\bar r & \bar r & \bar r \cr
                            \bar r & 1+\e+\bar r & \bar r \cr
                            \bar r & \bar r & 1+\d+\bar r \cr } \right)
	+ 2 \eta_R {c_\n} \left( \matrix{1 & 0 & 0 \cr
                            0 & 1 & 0 \cr  0 & 0 & 1 \cr } \right) \ , 
 \label{MassR}
 \end{equation}
\noindent
where
\begin{equation}
   \bar r = r - \frac{2}{3}\eta_R \ .
\end{equation}
\noindent
 Here we take $\eta_R\equiv \eta_e\simeq \eta_\mu$, which is a good approximation
 \cite{Haba}.  Its numerical value depends on  $\tan\b$ as:
 $10^{-2}$, $10^{-3}$ and $10^{-4}$ for $\tan \beta=60, \ 10,$ and $1$, respectively.
 As seen in eq.(\ref{Mass}) and eq.(\ref{MassR}),
   radiative corrections are absorbed into the original parameters
  $r$, $\e$ and $\d$ in the leading order.
 Thus the structure of the mass matrix is stable against radiative 
  corrections although our model leads to nearly degenerate neutrinos.
  
 Let us present  numerical results. 
 We take $L_{12}=-L_{21}=\sqrt{m_e/m_\mu}$ and
 $L_{23}=-L_{32}=-{m_\mu/m_\tau}$, which are suggested from the ones
 in the quark sector,  in eqs.(\ref{LMIX}).
 We show the result in the case of $\d=0.05$ 
  as a typical case.\footnote{Parameters $r$, $\e$ 
  and $\d$ are asuumed to be real. If they are taken 
  to be complex, the CP violation can be predicted as in ref.\cite{Xing}.}
 Putting $\Delta m_{\rm atm}^2=\Delta m_{32}^2=3\times 10^{-3}$ in eq.(\ref{dm2}),
 we get $c_\n=0.18 \eV$, which is consistent with the double beta decay experiment
 \cite{Double}.\footnote{
 The result is consistent with the constraint
 of the double beta decay experiment as far as $\d\geq 0.04$.}
 Taking $\e=0.002$ as a typical value,  predictions of 
 $\sin^2 2\th_{\rm atm}$ and $\sin^2 2\th_{\odot}$ are shown versus $r$ in fig.1.
 It is found that the predicted solar neutrino mixing lies in the region of
 the LMA-MSW solution if $r/\e=0.1\sim 0.5$ is taken, while the mixing of
 the atmospheric neutrino changes slowly. In this parameter region,
 $\Delta m_{\odot}^2= (1\sim 2)\times 10^{-4}\eV^2$ is predicted. 
 As far as $\d=\l^2\sim \l$ and $\e=\l^4\sim \l^3$,
 where $\l\simeq 0.22$,  obtained results
 are similar  to  the ones in fig.1.
  Thus the LMA-MSW solution with $\sin^2 2\th_{\rm atm}\geq 0.9$ is easily realized
 by taking a relevant $r/\e$ in this model.
 
  We have investigated  phenomenological implications on the neutrino
 flavor mixings in the $S_{3L}\times S_{3R}$ symmetric mass matrices
 including symmetry breaking terms. 
  We have shown how to get the LMA-MSW solution in this model.
  The non-zero value of the symmetric parameter $r$ is essential
   in order to get  $\sin^2 2\th_\odot=0.65\sim 0.97$.
   However, there is no reason that $r$ is very small in the $S_{3L}\times S_{3R}$ 
  symmetry, and so we need its extension, for example, 
    the $O_{3L}\times O_{3R}$ model \cite{O3},  which  leads to naturally 
    the small $r$ and
  the unique prediction of the LMA-MSW solution.
  It is found that radiative corrections are absorbed into the original parameters
  $r$, $\e$ and $\d$. Therefore, the structure of the mass matrix
  is stable against radiative corrections 
  although it leads to nearly degenerate neutrinos.
  Furtheremore, the neutrino mass matrix can be modified by introducing the CP 
  violating phase \cite{Xing}.
  We wait for results in KamLAND experiment \cite{KAMLAND}
  as well as new solar neutrino data.
	
  
 This research is  supported by the Grant-in-Aid for Science Research,
 Ministry of Education, Science and Culture, Japan(No.10640274).  
   
\newpage

\newpage

\begin{figure}
\epsfxsize=12 cm
\centerline{\epsfbox{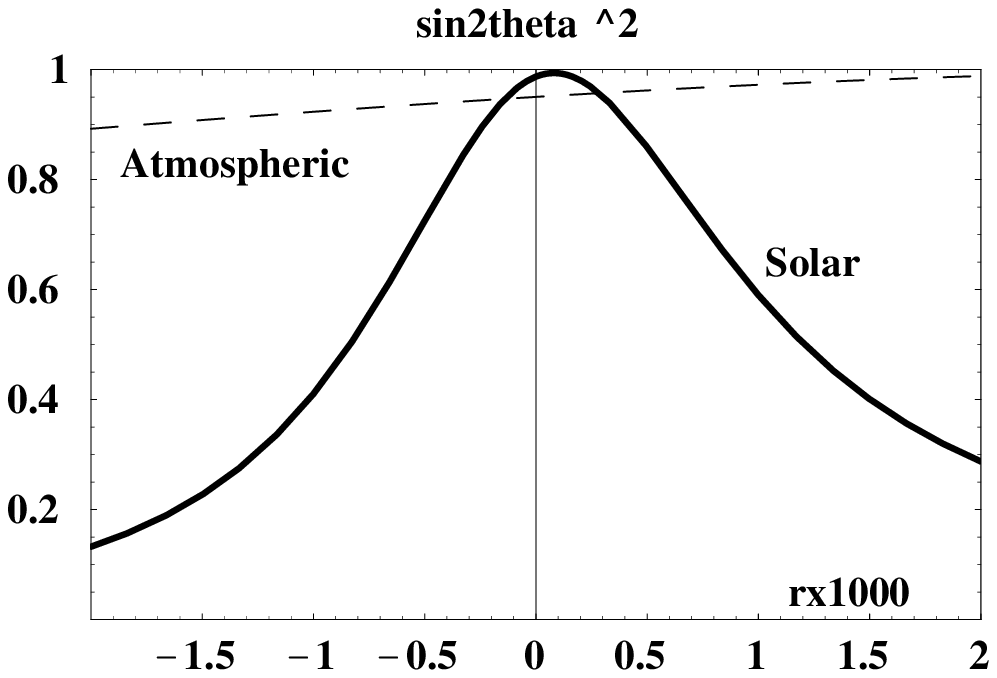}}
\caption{}
\end{figure}

Fig. 1: The $r$ dependence of  $\sin^2 2\th_{\rm atm}$ and $\sin^2 2\th_{\odot}$.
 $c_\n=0.18 \eV$, $\d=0.05$ and  $\e=0.002$ are taken. 
\end{document}